\documentclass[cscw]{acmart}

\AtBeginDocument{%
  }

\begin{document}

\title{Bridging Nodes and Narrative Flows: Identifying Intervention Targets for Disinformation on Telegram}

%
\author{Devang Shah}
\authornote{Both authors contributed equally to this research.}
\email{devangvshah16@gmail.com}
\author{Hriday Ranka}
\authornotemark[1]
\email{hridayr1234@gmail.com}
\affiliation{%
  \institution{SimPPL}
  \city{Mumbai}
  \state{Maharashtra}
  \country{India}
}

\author{Lynnette Hui Xian NG}
\affiliation{%
  \institution{Carnegie Mellon University}
  \city{Pittsburgh}
  \state{Pennsylvania}
  \country{USA}}

\author{Swapneel Mehta}
\affiliation{%
 \institution{SimPPL, Boston University and Massachusetts Institute of Technology}
 \city{Boston}
 \state{Massachusetts}
 \country{USA}}

%
\renewcommand{\shortauthors}{D. Shah and H. Ranka et al.}

\begin{abstract}
 In recent years, mass-broadcast messaging platforms like Telegram have gained prominence for both, serving as a harbor for private communication and enabling large-scale disinformation campaigns. The encrypted and networked nature of these platforms makes it challenging to identify intervention targets since most channels that promote misleading information are not originators of the message. In this work, we examine the structural mechanisms that facilitate the propagation of debunked misinformation on Telegram, focusing on the role of cross-community hubs—nodes that bridge otherwise isolated groups in  amplifying misinformation. We introduce a multi-dimensional ‘bridging’ metric to quantify the influence of nodal Telegram channels, exploring their role in reshaping network topology during key geopolitical events. By analyzing over 1,740 Telegram channels and applying network analysis we uncover the small subset of nodes, and identify patterns that are emblematic of information ‘flows’ on this platform. Our findings provide insights into the structural vulnerabilities of distributed platforms, offering practical suggestions for interventions to mitigate networked disinformation flows.
\end{abstract}

\ccsdesc[500]{Human-centered computing}
\ccsdesc[500]{Human-centered computing~Collaborative and social computing}
\ccsdesc[500]{Human-centered computing~Collaborative and social computing design and evaluation methods}
\ccsdesc[500]{Human-centered computing~Social network analysis}

\keywords{Social Network Analysis, Propaganda Networks, Platform Moderation, Distributed Messaging Ecosystems, Cross-Community Hubs, Misinformation Intervention }


\maketitle

\section{Introduction}
In the past decade, private messaging platforms have emerged as powerful vehicles for information dissemination, fundamentally altering the landscape of digital communication through the introduction of anonymity \cite{1, 2}. However, this transformation has brought with it unprecedented challenges, particularly in the realm of misinformation propagation. Telegram, with its encrypted channels and vast user base, has become a focal point for researchers and policymakers alike, as it represents a complex ecosystem where information—both accurate and misleading—can spread rapidly and with far-reaching consequences \cite{3Routledge, 4Jihad}.

The \textit{distributed}\footnote{As a platform, Telegram operates on a centralized computing systems architecture but its channel-based structure \textit{decentralizes the information feed for users}, and for the purpose of this research, this lack of a central information proliferation is what the term \textit{distributed} will refer to.}\cite{ng2024} nature of information proliferation on Telegram, characterized by interconnected channels and groups, creates an environment ripe for the formation of echo chambers and information silos\cite{5Törnberg, 6Pushshift}. 

Within this intricate network structure, certain nodes play a pivotal role in bridging disparate communities, acting as conduits for information flow across ideological and thematic boundaries. These "bridge nodes" are gateways for both, the dissemination of reliable information and the amplification of misinformation
\cite{7Urman}.

Past research has advanced our understanding of misinformation dynamics in social media platforms \cite{8Lynnette, 9mehta2022estimating, 10Matteo}. This includes content-based analyses \cite{FAN2020101498} and network metrics to identify influential nodes and information flow patterns. Building on this foundation, our research examines the structure of private messaging ecosystems like Telegram, where the dynamics of information propagation may differ significantly from centralized ‘feed-based’ platforms, posing unique risks \cite{1}.
Our research takes a multidimensional approach in identifying and analyzing the role of critical nodes in Telegram's network structure. We propose a "bridge score" metric that aggregates measures across varied network characteristics to provide a comprehensive understanding of each node's potential to act as an ‘information hub’. This approach builds upon traditional graph-centrality measures, offering new insights into the structural underpinnings of information flow in distributed messaging platforms. 

\section{Key Contributions}
In this research, we seek to answer two key questions:

\begin{enumerate}
    \item Are there distinct communities within which debunked disinformation propagates on distributed messaging ecosystems like Telegram?
    \item How consistent is a multi-dimensional bridging metric in measuring the contribution of cross-community hubs in amplifying misinformation and reshaping network topologies?
\end{enumerate}

We validate the metric using data from 1,748 public Telegram channels and groups containing over 900,000 messages, using it to generate a minimum viable set of targets for intervening on the flow of disinformation across this large network. Our contributions are twofold: 

\subsection{Examining the nature of Fact-checked misinformation propagation across Communities} 
We collect links to debunked disinformation from a third-party website, EUvsDisinfo.eu, a flagship project of the European Union that is focused on fact-checking false claims transparently.  We identify messages containing these links shared in Telegram channels and  visualize its spread to construct a misinformation network of channels that interact to amplify the false claims. We split the network into separate communities based on their primary narratives. We observe that misinformation crossing community boundaries (defined by modularity classes) reached an average of 3.1 times more views than misinformation contained within a single community. Interestingly, the communities that are primarily utilized to forward messages and amplify misinformation have a 57.5\% higher engagement rate than the community where the false narrative originates. Much of the literature on channels where pro-Russian disinformation is promoted focus on these channels, where even removals may not have a long-term effect since the originating nodes are unaffected. These findings underscore the role of modularity in network structure in shaping the spread and impact of misinformation on Telegram.

\subsection{Developing a Multi-Dimensional Bridging Metric to identify targets for intervention} 
Our multi-dimensional bridging metric, which combines in-degree centrality, eigenvector centrality, and clustering coefficient, revealed crucial insights into the role of cross-community hubs. Our selected bridge metric identified 12 (\textit{originally 14}\footnote{Further research will focus on the 12 publicly accessible channels, as data from private groups was unavailable for analysis.}, but 2 of these were private communities not accessible to the general public) high-impact bridge nodes across the Telegram network, which were responsible for major cross-community misinformation flows. Removal of the top 12 bridge nodes (by our metric) significantly impacted network topology and resulted in a 33.33\% rise in the number of communities. We also observed that the Russian disinformation machinery reuses the same Telegram channels to promote different campaigns like Russia-Ukraine conflict, Moscow (Crocus Hall) attack, Anti-West propaganda, etc. 
Analysis of temporal dynamics revealed that the influence of bridge nodes on misinformation spread increased by 24\% during periods of heightened global events (e.g., feud between Russian president and Wagner group, pandemic peaks, Crocus Hall attack), suggesting their crucial role in misinformation dissemination during critical times.
These insights contribute to the broader understanding of information propagation in distributed systems, offering valuable knowledge that extends beyond Telegram to other messaging platforms and social networks \cite{zannettou2017webcentipedeunderstandingweb}. Our work provides a foundation for developing more effective strategies to stop disinformation proliferation \& promote information integrity in these complex digital environments.
As society grapples with the challenges of misinformation and the erosion of shared truths \cite{Starbird}, our research offers a crucial step towards understanding and potentially mitigating these issues in distributed messaging platforms. By illuminating the structural underpinnings of information flow in these ecosystems, we pave the way for more targeted and effective interventions in the ongoing battle against digital misinformation.

\section{Related Work}
Telegram presented itself as a lightly moderated platform, with little to no government oversight; an image that has recently been questioned given its self-admitted compliance with the governments of India and Brazil\footnote{\texttt{\url{https://t.me/durov/346}}} Following the ban of many Russian news outlets throughout Europe, several have turned to Telegram to share their content, with Ukraina, Russia Today and Sputnik News even dedicating pages to instruct users on how to download the app (Bovet and Grindrod 2022 \cite{14Bovet}). Telegram serves as a medium for information propagation, with its focus on several critical themes: the interaction between Telegram and media narratives, the coordinated manipulation by bot/fake accounts, contextual factors influencing user behavior, and the challenges against effective content moderation strategies.\\
\textbf{Information Dissemination Dynamics.} Hanley and Durumeric (2023) \cite{15Hanley} analyzed the interrelation between Russian media outlets and 732 Telegram channels over a period of 1 year. They identified Telegram as a key source of content for Russian media outlets, some of which are using discussions from Telegram  as a citation for the origin of the information, for up to 26.7\% of their articles. Hoseini et al. (2022) \cite{16Hoseini} further conducted an in-depth analysis of over 140 million messages coming from over 9,000 public Telegram channels. They established that a small number of users, whom they termed "superspreaders," are responsible for a  particularly disproportionate amount of messages. Their work exhibits the affordances that underpin the delivery of content on Telegram; although the platform offers immense accessibility and very extensive reach, delivery of content seems relatively localized. This phenomenon mirrors similar research in which just twelve prominent individuals, known as the "Disinformation Dozen" \footnote{\texttt{\url{https://counterhate.com/research/the-disinformation-dozen/}}}, were found to account for nearly two-thirds of anti-vaccine content circulating on major social media platforms, underscoring how a small group can disproportionately influence information dissemination across networks.\\
\textbf{Coordinated Information Manipulation.} Studies have also revealed the strategic manipulation of information by inauthentic accounts. Burghardt et al. (2022) \cite{17Burghardt}  investigated Russian-affiliated accounts' coordinated efforts during the 2017 French election, showing how these accounts amplified certain narratives through repetitive retweets and emotionally charged content. This coordination is also reflected in the work of Dash and Mitra (2024) \cite{18Dash}, who studied Indian Twitter’s influence campaigns, highlighting the need to draw a distinction between"disseminators" and "amplifiers" within hashtag campaigns. Both studies reveal  the complex tactics used to shape and manipulate online conversations, relevant for understanding how similar mechanisms might work in Telegram's lightly moderated  ecosystem.\\ \\
\textbf{Contextual Factors in Information Practices.} Beyond manipulation, contextual influences strongly shape how users interact with misinformation on Telegram. Nikkhah et al. (2021) \cite{19Nikkhah} emphasized how Iranian immigrants largely used Telegram as an essential source for immigration-related information, indicating how the functionalities of the platform inform particular information-seeking behaviors. Peeters and Willaert (2023) \cite{20Peeters_Willaert_2022} have further explored the role of Telegram in the diffusion of conspiracy theories, and elaborated on how it serves as a platform through which interlinked communities promote conspiracy-theories via message-forwarding. Lim and Perrault (2023) \cite{21Lim} found that in Singapore, proliferation of misinformation on telegram is best explained by sharing behavior rather than the production of new content; this further contributes toward the argument that user activities, particularly sharing behavior, are significant in shaping the way information spreads.\\
\textbf{Content Moderation and Platform Governance.} Ma (2023)\cite{22Ma} not only advances the case for involving content creators in the design of moderation systems but also details socio-economic dimensions of current policies and bureaucratic barriers to accessing the system. Participatory design initiatives, therefore, are needed to balance the interests of different stakeholders involved in moderation efforts. Adjusting moderation strategies to the idiosyncrasies of Telegram could support a healthier information ecosystem on the platform and facilitate better governance.

Several key studies from the community inform our understanding of how disinformation circulates in distributed and semi-regulated ecosystems like Telegram. Hanley et al. \cite{hanley2023goldenageconspiracytheories} explored the interrelationships among conspiracy theories across five domains (including QAnon, COVID, UFOs, among others) that are interconnected by both legitimate news sources and misinformation platforms. This research emphasizes the role of hyperlinked networks in propagating false narratives, drawing attention to the intricate web of associations among misinformation centers. This is relevant to our focus on Telegram, where disinformation can spread across seemingly isolated channels, supported by similar interconnected dynamics. Similarly, Nied et al. \cite{24Nied} explored networks of alternative crisis narratives. Their identification of automated accounts and social botnets' role in the spread of disinformation demonstrates how these automated systems can rapidly amplify false narratives across multiple communities, making traditional fact-checking efforts ineffective due to the speed and scale of propagation. Understanding community structure and automated orchestration is crucial because it allows platform moderators to identify and disrupt these amplification networks before they can achieve widespread reach, and helps researchers develop more effective early warning systems for emerging disinformation campaigns. Aghajari’s \cite{25Aghajari} work moves beyond individual content analysis to focus on community-level impacts. This ecological view aligns with our research goal of analyzing not only how misinformation propagates but also how certain nodes and hubs within Telegram’s network structure serve as amplifiers, shaping community interactions and influence.
In this way, our work offers  a comprehensive methodology supported by empirical evidence at a large scale, for understanding how distributed messaging platforms reshape disinformation ecosystems, with implications for content moderation and platform governance.

\section{Dataset}
\begin{figure}[h]
  \centering
  \includegraphics[width=0.9\linewidth]{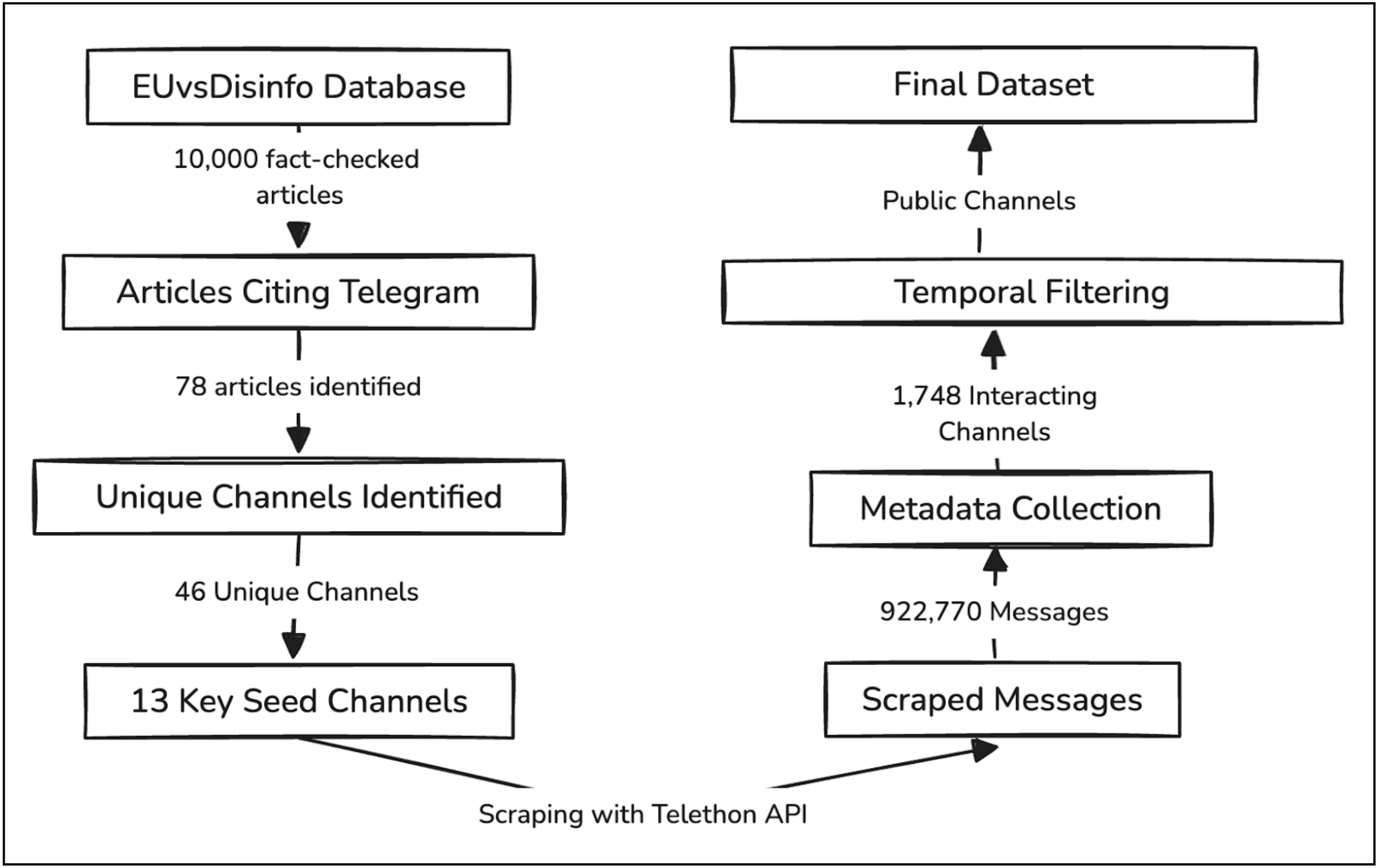}
  \caption{Architecture diagram of the data collection pipeline}
  \Description{Data collection pipeline}
\end{figure}
In this section, we describe our data collection and processing pipelines. Figure 1 illustrates our system.

\begin{table}
  \label{tab:commands}
  \begin{tabular}{ccccc}
    \toprule
    Channel Type &Channels &Subscribers &Number of Messages  &Number of Forwarded messages\\
    \midrule
    Seed Channels &13 &5,009,485 &922,770  &210,534\\
    Bridge Channels &12 &  5,153,011 &125,266 &99,336\\
    \midrule
    Total &25 & 10,162,496 &1,048,036 &309,870\\
    \bottomrule
  \end{tabular}
  \caption{Overview of our Telegram Dataset}
\end{table}

We collected messages from public Telegram channels using a multi-stage process, combining authoritative disinformation databases with targeted channel data mining. Our seed article data source was the EUvsDisinfo database\footnote{\href{https://euvsdisinfo.eu/}{https://euvsdisinfo.eu/}}, maintained by the European External Action Service's East StratCom Task Force. We extracted over 10,000 fact-checked articles from this database. Analysis of the EUvsDisinfo dataset revealed 78 articles specifically citing Telegram as the source platform, leading to the identification of 46 unique Telegram channels. To focus on the most influential actors, we used TGSTAT\footnote{\href{https://tgstat.com/}{https://tgstat.com/}}, a Telegram analytics platform, to rank these channels by their citation index. This process yielded a subset of 13 key channels (referred to as "seed channels") with the highest potential for disseminating and amplifying disinformation narratives.

Using the Python Telethon API\footnote{\href{https://docs.telethon.dev/en/stable/}{https://docs.telethon.dev/en/stable/}}, we scraped posts from these 13 seed channels. Our data collection resulted in two primary datasets: one containing metadata about 1,748 Telegram channels that interacted with the 13 seed channels, and another comprising 922,770 messages from those seed channels. The resulting dataset is predominantly in Russian and English languages. To focus on the Russian invasion of Ukraine, we applied temporal filtering to isolate 2421 forwards that occurred after January 2022.

In our study, we exclusively used data from publicly available channels and did not seek out private channels or conversations. Furthermore, we focused solely on a user’s interactions visible in public channels and did not seek to identify individuals beyond their engagement during the given period, in keeping with general privacy guidelines for collecting social media.

\section{Methodology}

\subsection{Network Graph Creation}
Identification of hub nodes begins with the construction of a comprehensive network graph that visually represents the flow of information across Telegram channels. This graph serves as the foundation for subsequent analyses, providing crucial insights into the structure and dynamics of disinformation dissemination.

The network graph construction initiates with the identification of forwarded messages within the collected dataset. Each unique Telegram channel is represented as a node in the graph. A directed edge is created between two nodes when a message is forwarded from one channel to another, representing the flow of information.
Drawing inspiration from the telegram-tracker project by Esteban Ponce de Leon \footnote{\href{https://github.com/estebanpdl/telegram-tracker}{https://github.com/estebanpdl/telegram-tracker}}, a custom Python script to process the msgs\_dataset.csv \& collected\_chats.csv files was developed. This script performs the following key operations:

\begin{itemize}
    \item \textbf{Iteration through outgoing messages}: The script iterates through each entry in the collected\_chats\.csv file, examining the `source` \& `username` fields to identify forwarded messages outgoing from seed channels.
    \item \textbf{Iteration through incoming messages}: The script iterates through each entry in the msgs\_dataset\.csv file, examining `forward\_msg\_from\_peer\_name` \&  `channel\_name` fields to identify forwarded messages incoming into seed channels.
    \item \textbf{Node creation}: For each unique channel encountered (both source and destination of forwards), a node is created in the graph if it doesn't already exist.
    \item \textbf{Edge creation}: When a forwarded message is identified, an edge is created from the source to destination channel.
\end{itemize}

The resulting graph structure is stored using the NetworkX library, which provides a flexible and powerful framework for network analysis. The NetworkX graph data is saved as a GEXF (Graph Exchange XML Format) file. 

\subsection{Community detection}
After constructing the network graph, we identify cohesive communities within the graph and analyze the thematic content of these communities. 
The Louvain algorithm \cite{26Blondel}, a widely-used method for detecting communities in large networks, was employed to identify modular classes within the Telegram channel network. This algorithm is particularly effective for uncovering hierarchical community structures in complex networks.

\begin{enumerate}
    \item \textbf{Data Preparation:} Data Preparation: The GEXF file generated from the network graph creation step was loaded into Gephi, an open-source network analysis and visualization software.
    \item \textbf{Algorithm Application:} Algorithm Application: Within Gephi, the Louvain community detection algorithm was executed on the loaded graph. The resolution parameter, which controls the granularity of the detected communities, was set to 2.2 after careful consideration and experimentation. This value was chosen to strike a balance between detecting meaningful communities and avoiding over-fragmentation of the network.
    \item \textbf{Execution and Output:} Execution and Output: The algorithm iteratively optimized the modularity of the network partition, grouping nodes into communities that maximize internal connections while minimizing external connections.
\end{enumerate}

The application of the Louvain algorithm resulted in the identification of 6 distinct modular classes within the network. These classes were visually distinguished in the Gephi visualization using a color scheme: [Green, Purple, Dark Green, Orange, Red, Blue]. Each color represents a distinct community of Telegram channels that exhibit higher internal connectivity compared to their connections with channels in other communities. Figure 2 displays the distribution of community sizes, revealing a notable disparity. The largest community encompasses 30.72\% of all nodes, while the smallest consists of 4.06\% of the network.
Gephi also offers a bunch of different layouts, determining the nature of organizing the graph visually. We utilised the ‘ForceAtlas’ layout, since this algorithm pulls strongly connected nodes together and pushes weakly connected nodes apart. Its complexity is O(N²). This step transforms the raw message data into a structured network representation.

\begin{figure}[h]
  \centering
  \includegraphics[width=0.80\linewidth]{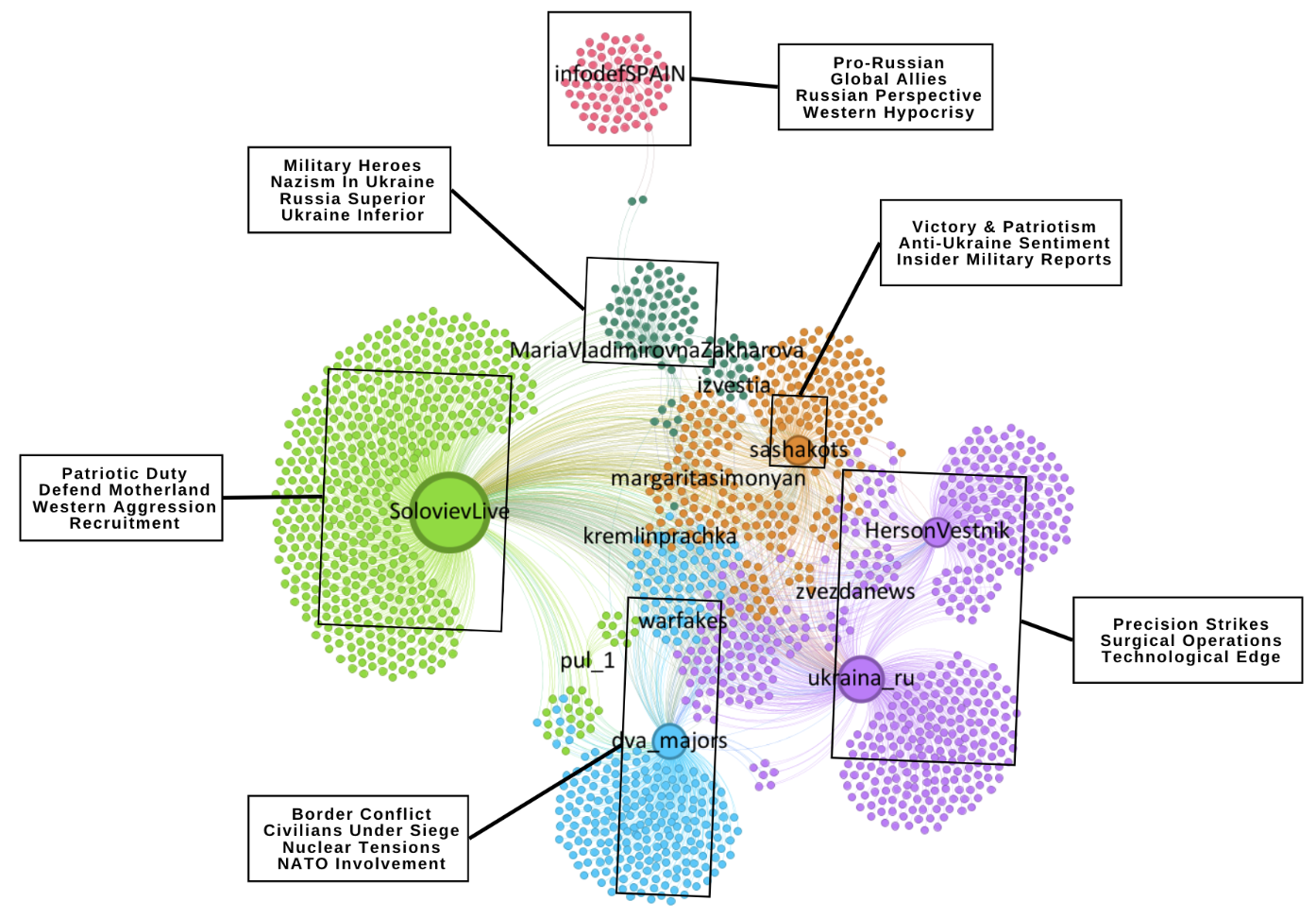}
  \caption{Visualization of Telegram Misinformation network on Gephi}
  \Description{Network Visualization on Gephi}
\end{figure}

\subsection{Bridge Metric Formulation}
Building upon the analysis of the above Telegram channels and their message sharing network structure, this study identified six distinct communities propagating diverse forms of misinformation. While these communities exhibited strong internal cohesion, the overall network analysis revealed a more complex picture of information flow and influence. Despite the apparent isolation of these communities, certain nodes emerged as critical connectors, facilitating the spread of information across community boundaries. The observation of these inter-community connections prompted a deeper investigation into the role of bridging nodes or hubs within the network. These nodes appeared to play a disproportionately significant role in the dissemination of misinformation across the broader network. Their position at the intersection of multiple communities suggested a unique capacity to amplify and cross-pollinate narratives, potentially accelerating the spread of misinformation beyond isolated echo chambers.
This work introduces a novel multi-dimensional bridging metric designed to identify and quantify the impact of cross-community hubs in distributed messaging ecosystems, particularly in the context of misinformation amplification. The proposed metric, termed the Bridge Score, integrates three key network centrality measures: in-degree centrality, eigenvector centrality, and clustering coefficient. Combining these three allows us to identify nodes with extensive connections, strategic positioning, and crucial bridging roles respectively. This composite approach aims to capture the multifaceted nature of influential nodes within complex network structures. The Bridge Score is formulated as a weighted sum of these normalized centrality measures, allowing for flexible parameterization to adapt to various network contexts and research objectives.
The Bridge Score is crucial for identifying nodes that not only have high connectivity but also serve as key connectors between different communities in a network. By integrating local and global network properties, this metric enhances our understanding of a node's role in spreading information and maintaining network cohesion, which is especially useful in analyzing misinformation. This approach provides a more comprehensive assessment of node importance than single-metric analyses and reveals structural vulnerabilities and resilience by observing changes in network topology when high-scoring nodes are removed. Beyond misinformation studies, the Bridge Score's flexibility makes it valuable for various network analysis applications, contributing a composite metric  relating to graph theory. We begin our discussion with each individual metric contributing to the composite, examining how indegree centrality, eigenvector centrality, and clustering coefficient each play a unique role in shaping the Bridge Score.

\subsubsection{In-degree Centrality - Quantifying Direction:}
In-degree centrality is a vital component of the Bridge Score, offering a direct measure of a node's prominence within the network. In the context of directed networks, such as those formed by Telegram channels and their message sharing, in-degree centrality quantifies the number of incoming edges to a given node. Following the foundational work of Freeman \cite{FREEMAN1978215} and as elaborated in modern network theory by Newman \cite{28Newman}, we define the in-degree centrality of a node (channel) as the number of incoming edges (forwarded messages) into it.
This metric is crucial for identifying potential bridge nodes for several reasons:
\begin{itemize}
\item{Information Reception: Nodes with high in-degree centrality are positioned to receive information from multiple sources, making them potential aggregators of diverse narratives and misinformation strains.}
\item {Influence: In the context of social networks, a high in-degree often correlates with greater visibility and perceived authority, factors that can amplify the spread of information or misinformation.}
\item{Community Interface: Nodes with high in-degree centrality that accept connections from multiple communities are uniquely positioned to serve as conduits for cross-community information flow.}
\end{itemize}

The inclusion of in-degree centrality in the Bridge Score calculation addresses a critical aspect of information dissemination in distributed messaging ecosystems. While a high in-degree alone does not necessarily indicate a bridging role, it provides a crucial foundation for identifying nodes that have the potential to do so. A high in-degree may indicate not just popularity, but also vulnerability to diverse information inputs, including misinformation from multiple sources (as observed in the case of coordinated campaigns, or in denial of service attacks).

\subsubsection{Eigenvector Centrality - Capturing Influence:}
Eigenvector centrality extends the concept of node importance by considering not only the quantity of connections but also the quality of those connections within the network structure. This metric captures the idea that connections to people who are themselves influential will lend a person more influence than connections to less influential people.

Newman \cite{28Newman} recognizes that connections to highly influential nodes contribute more to a node's importance than connections to peripheral nodes. This property is particularly relevant in the context of information dissemination and misinformation amplification in distributed messaging ecosystems. 
Incorporating eigenvector centrality into our bridge score formulation models influence propagation by highlighting nodes that are influential, and therefore well-positioned to spread misinformation across various communities. This metric also emphasizes strategic positioning, identifying nodes that, while not having the highest number of connections, are relevant given to their connections with other influential nodes, revealing potential ``hidden influencers`` in misinformation networks. Assuming reliable information receives more engagement, the robustness of eigenvector centrality to manipulation is a key advantage; it is less susceptible to the artificial inflation of influence through numerous low-quality connections, making it a more reliable measure of genuine influence. Lastly, the recursive nature of eigenvector centrality captures the potential for cascade effects in information diffusion, a conceptual underpinning of virality.
By combining these metrics, we create a measure that accounts for both the direct connectedness of a node and its strategic positioning within the broader network structure.

\subsubsection{Clustering Coefficient - Measuring Community Connectivity:}
The local clustering coefficient is a metric that captures the extent to which a node's neighbors are connected to each other, providing insights into the formation of local network structures and communities. In network theory, the clustering coefficient is closely associated with the concept of network modularity, which quantifies the tendency of a network to organize into distinct communities or clusters. Nodes with a high clustering coefficient are typically embedded within dense clusters or communities, while nodes with a low clustering coefficient are more likely to act as bridges connecting different communities.

Nodes with low clustering coefficients serve as critical bridges between densely connected communities and support information flow between distinct groups. Such nodes often occupy structural gaps in the network, acting as key intermediaries that control information flow. These Telegram channels also function as information brokers, selectively transmitting or withholding information, which can contribute to the formation of echo chambers and divergent narratives. In fact, removing these key nodes, identified by their low clustering coefficients, can lead to significant network fragmentation, impairing the efficient flow of information and restricting the spread of misinformation. The incorporation of the clustering coefficient into our bridge score metric allows us to capture the community dynamics and local network structures.

\subsection{Weight Optimization for Composite Scoring:} 
By combining these three metrics – in-degree centrality, eigenvector centrality, and clustering coefficient – our bridge score formulation offers assessment of a node's potential to serve as a cross-community hub, facilitating the amplification of diverse narratives and reshaping network topologies. 
To ensure comparability and balanced integration of these metrics, we employ a normalization process followed by a weighted sum approach. Each metric is first normalized using min-max normalization to scale values between 0 and 1.
The final bridge score (BS) for each node is then calculated as a weighted sum of these normalized metrics:
\begin{equation}
\text{Bridge\_Score} = (w_i \cdot \text{indegree\_centrality}) + (w_e \cdot \text{eigenvector\_centrality}) + (w_c \cdot \text{clustering\_coefficient})
\end{equation}
where w\textsubscript{i}, w\textsubscript{e}, w\textsubscript{c} are respective weights for each metric.

This formulation allows for flexible adjustment of the relative importance of each metric through the weight parameters, enabling fine-tuning of the bridge score calculation to best capture the dynamics of specific messaging ecosystems under study.
To maximize the effectiveness of our bridge score metric in identifying influential cross-community hubs, we implemented an iterative weight optimization process. This process aims to determine the optimal combination of weights for in-degree centrality, eigenvector centrality, and clustering coefficient that best captures the most influential bridging nodes. 
The empirical process involved systematically varying the weights (w\_i, w\_e, w\_c) from 1 to 10 in integer increments each, resulting in 1000 unique weight combinations. This exploration allows us to capture subtle variations in the relative importance of each metric. Beyond 10, we find it is computationally expensive to calculate this score, while below 1 it is too sensitive to small changes so we believe a reasonable set of weights can be discovered to fall within this range. For each weight combination (w\_i, w\_e, w\_c), we performed the following steps:

\begin{itemize}
    \item {\itshape Bridge Score Calculation:} We calculated the bridge score for each node in the network using the formula in equation 1.
\item {\itshape Identification of Top Bridge Nodes:} For each weight combination, we identified the top 10 nodes with the highest bridge scores.
\item {\itshape Network Fragmentation Analysis:} To assess the impact of these top bridge nodes on the overall network structure, we performed a network fragmentation analysis. This involved (a) Creating a copy of the original network to preserve the baseline structure, (b) Removing the top 12 bridge nodes from the copied network, (c) Calculating network-level metrics for both the original and modified networks. We focused primarily on network density as our key metric for this analysis, as it provides a concise measure of overall network connectivity. Network density is calculated as
D = 2 * |E| / (|V| * (|V| - 1))	
where, |E| is the number of edges and |V| is the number of vertices in the network.
\item {\itshape Impact Quantification:} We quantified the impact of removing the top bridge nodes by calculating the difference in network density:
\begin{equation} 
\Delta_{\text{density}} = D_{\text{original}} - D_{\text{modified}}
\end{equation}
A larger \begin{math} \Delta_{\text{density}} \end{math} indicates a more significant disruption to the network structure, suggesting that the removed nodes played a crucial role in maintaining network connectivity and information flow.
\item {\itshape Comparative Analysis:} We compared the \begin{math} \Delta_{\text{density}} \end{math} values across all weight combinations to identify those that resulted in the most substantial network disruption when top bridge nodes were removed. This method helps uncover non-linear relationships between metric weights and their effects on network structure. As highlighted by \cite{29Coscia}, the exploration of different parameter settings can significantly affect the resulting network topology. We emphasize network density as a key metric, which reflects overall connectivity and information flow within the network. A notable reduction in density following node removal suggests those nodes were crucial for maintaining network cohesion, in line with the structural holes theory \cite{30Burt}. Our focus on the top bridge nodes is informed by network resilience principles, recognizing that a few highly connected nodes can disproportionately influence network structure and efficiency \cite{Albert_2000}. Additionally, we employ perturbation analysis to assess node importance by evaluating their impact on network dynamics when altered or removed, drawing from minimum driver node set concept in controllability of complex networks. The outcome of these iterations are mentioned in section 6 below. 
\end{itemize} 

\section{RQ1: We can identify distinctive patterns in Fact-checked Misinformation propagation across Telegram}
Now, we delve into the spread of authority-identified misinformation across Telegram’s network structure, examining how different network characteristics influence propagation patterns. By exploring the interplay between community detection and topic modeling, our analysis uncovers how misinformation flows between distinct groups within the network.

\subsection{Setup:} After detecting communities in the authority-identified misinformation network using the Louvain Clustering Algorithm, a topic modeling approach was employed to extract the primary themes and narratives within each community. The BERTopic model, a state-of-the-art topic modeling technique based on transformers, was utilized for this purpose. Specifically, the "paraphrase-MiniLM-L12-v2" embedding model was selected since it captures semantic nuances in short texts, making it particularly suitable for analyzing Telegram messages.
The process involved aggregating all messages from the seed channels present in each identified community. These aggregated message sets were then fed into the BERTopic model, which leveraged its underlying BERT architecture to generate contextualized embeddings for each message. The model then applied c-TF-IDF to create dense clusters of semantically similar messages, effectively identifying the most prominent topics within each community. To ensure optimal performance and account for language diversity, all messages were translated to English using Google Translate \footnote{\href{https://translate.google.com/}{https://translate.google.com/}}.
The output of this analysis revealed distinct primary narratives for each color-coded community, as displayed in Figure 2. The figure delineates the community color, the seed channels included in each community, and a concise description of the primary narrative or theme identified by the BERTopic model.This approach not only quantifies the thematic focus of each community but also provides valuable insights into the specialized nature of disinformation dissemination across different channel clusters. The results underscore the efficacy of combining network analysis with advanced natural language processing techniques in unraveling the complex landscape of coordinated disinformation campaigns on encrypted messaging platforms.

\subsubsection{Thematic Analysis of Misinformation Across Communities} The analysis of the BERTopic model across the 6 communities in Figure 2 revealed distinct thematic focuses. The \textbf{Blue} community, represented by channels such as @dva\_majors, @warfakes, and @kremlinprachka, predominantly discusses military activities and border tensions involving Ukraine, Russia, and NATO. It often disseminates pro-Russian messages, including false reports about attacks on civilian infrastructure like hospitals and maternity wards, alongside narratives concerning Russian military mobilization and nuclear concerns. The \textbf{Green} community, including channels like @SolovievLive and @pul\_1, promotes Russian military and political propaganda, advocating for support of Russia's invasion of Ukraine, recruitment for the Wagner PMC, and framing Russia and Belarus as defenders against Western aggression. 
This community emphasizes themes of Russian military strength, demonizes Ukraine and the West, and encourages patriotic duty, often tying these narratives to Orthodox Christian values. The \textbf{Dark Green} community, represented by channels such as @izvestia and @MariaVladimirovnaZakharova, reinforces narratives of Russian military superiority and success in Ukraine while depicting Ukrainian forces as weak or cruel. The messages highlight Russian military achievements and frequently accuse Ukrainian forces of war crimes, portraying Russia as a protector of civilians in contested regions. In contrast, the \textbf{Red} community, exemplified by @infodefSPAIN, offers a pro-Russian perspective that critiques Western narratives and discusses Russia's geopolitical relations, including its military operations and ties with regions like Africa and Latin America. The \textbf{Purple} community, with channels such as @ukraina\_ru, @HersonVestnik, and @zvezdanews, glorifies Russian precision strikes against Ukrainian military assets, emphasizing successful attacks and portraying them as strategically significant while undermining Ukrainian capabilities. Finally, the \textbf{Orange} community, represented by @sashakots and @margaritasimonyan, communicates narratives of Russian military successes and Ukrainian failures, downplaying setbacks and spreading disinformation about alleged Ukrainian atrocities. This community aims to rally domestic support for the war effort by fostering a tone of Russian patriotism and providing purported insider military information.

\subsection{Network Scale, Connectivity and Topology:} Our analysis of the Telegram messaging ecosystem revealed a complex and interconnected network structure, providing insights into the potential pathways for misinformation propagation. The examined network comprised 1,748 nodes (Telegram channels) connected by 2,421 edges, representing interactions or information flows between these channels. This substantial scale underscores the potential for rapid information dissemination within the ecosystem.
Figure 2 presents a high-level visualization of the network, with nodes colored by the 6 detected communities. This visualization highlights the network's non-uniform structure, characterized by dense clusters interconnected by bridging connections. This topology indicates the presence of highly connected hub nodes, which play a crucial role in information dissemination and potentially serve as amplifiers of misinformation.


A few key network statistics:
Average path length is a global metric that represents the average number of steps required to connect one channel to another across the network. In Telegram’s ecosystem, a low average path length indicates that information reaches distant parts of the network quickly, even across ideologically or geographically distinct communities.
Network Density quantifies how many actual connections exist relative to the total possible connections among channels. A higher density means that channels are more interconnected through frequent message forwarding, allowing information to spread rapidly across the network.\\
Average path length: 1.579 \\
Network Density: 0.0015849 \\
The relatively short average path length, despite the network's large size, suggests a "small-world" property \cite{watts_collective_1998}, facilitating rapid information spread across the network.
The network structure with the presence of a scale-free topology, coupled with a clear community structure and geographical concentrations, suggests that misinformation propagation is likely to follow non-uniform patterns, potentially amplified by key hub nodes and bridge connections between communities. 

\section{RQ2: Combining In-Degree, Clustering Coefficient, and Eigenvector Centrality serves as a robust measure to identify a small set of targets for intervention}
Here we will explore the multi-dimensional bridging metric, which is pivotal in understanding how cross-community hubs influence the spread of misinformation and alter network topologies. By examining these bridge nodes, we will gain insight into their strategic positioning within the network and their role in connecting otherwise isolated communities. This analysis is crucial for identifying key points of vulnerability where misinformation is amplified, particularly during geopolitical events. Furthermore, the removal of these nodes reveals significant shifts in network metrics and engagement patterns, highlighting their impact on the overall flow of information. This exploration helps deepen our understanding of how distributed messaging ecosystems are shaped by influential connectors.

\subsection{Weight Optimization and Network Perturbation Analysis:}
The weight optimization process coupled with network perturbation analysis allowed us to fine-tune our metric while simultaneously assessing its ability to identify nodes crucial for maintaining network cohesion and information flow. 
For the weight optimization process, 1,000 weight combinations for bridge score components were assessed by calculating bridge scores for each combination, identifying the top 12 nodes, and performing network perturbation analysis. This analysis involved comparing the network density of the original and perturbed networks to measure the impact of removing the top 12 bridge nodes and finding the most optimal weights.

\begin{figure}[h]
  \centering
  \includegraphics[width=0.75\linewidth]{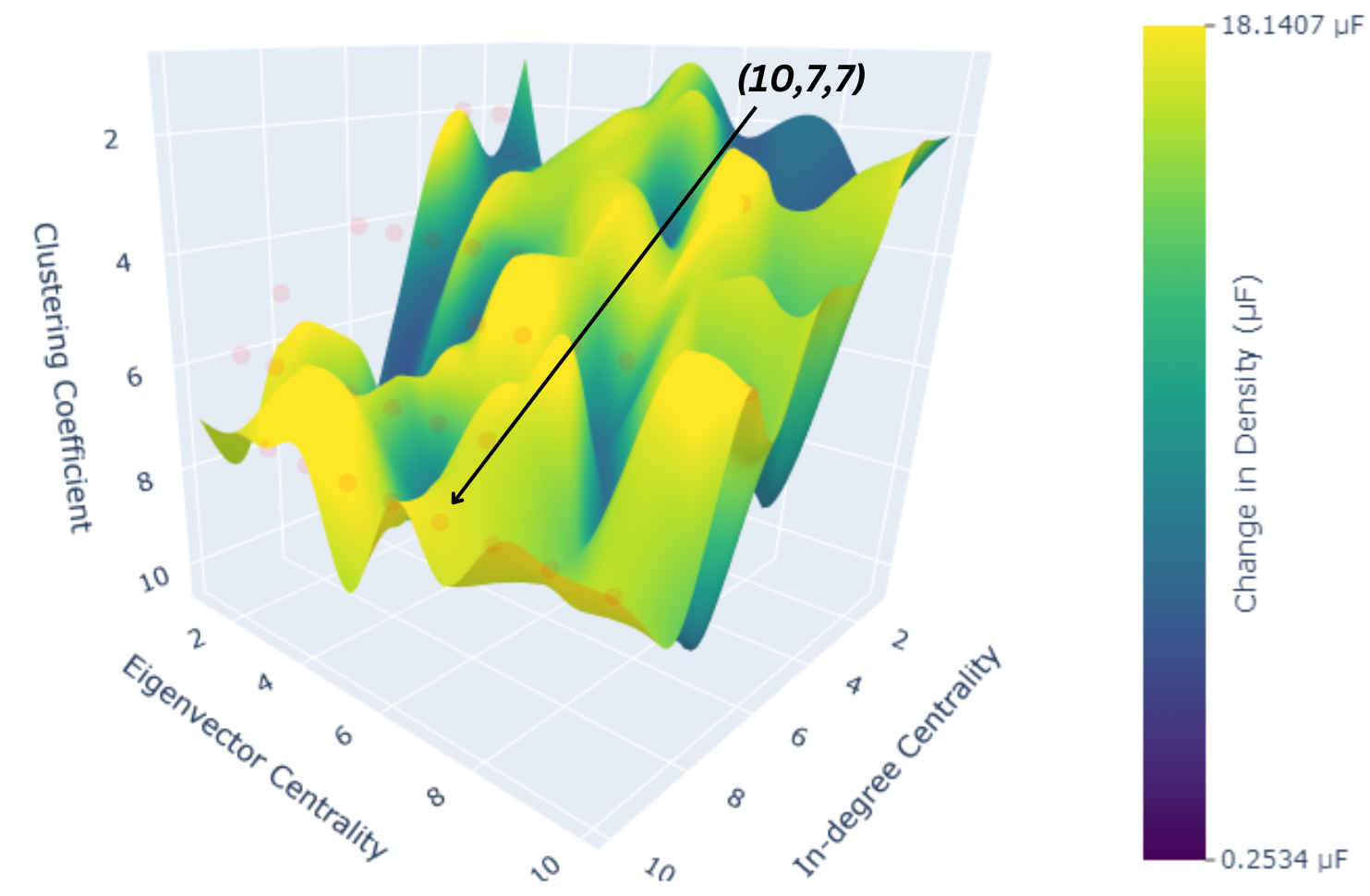}
  \caption{Weight combinations and corresponding change in global network density}
  \Description{3d plot}
\end{figure}

\begin{figure}[h]
  \centering
  \includegraphics[width=0.85\linewidth]{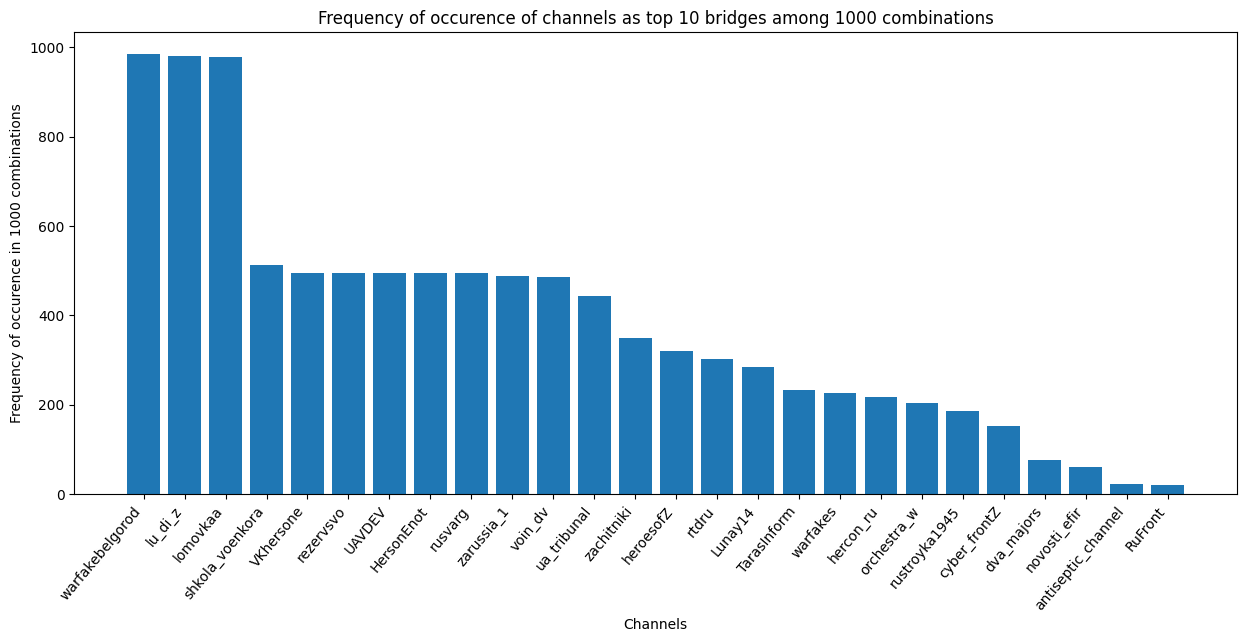}
  \caption{Frequency of occurrence of channel as a top bridging channel among 1000 weight combinations}
  \Description{histogram}
\end{figure}

Figure 3 demonstrates various weight combinations of Eigenvector Centrality, Clustering Coefficient, and In-degree that achieve significant reductions in network density by removing the top bridge channels. Importantly, these different combinations consistently identify the same core set of bridge nodes, each of which appears frequently among the top 12 most impactful cross-community hubs for the considered 1000 combinations (Figure 4). This recurrence underscores that our goal is not to seek a single “most optimal” weight configuration but rather to leverage our bridge metric to reliably uncover influential bridge nodes that act as key conduits of misinformation across communities. Regardless of the chosen optimal weight combination, the proposed bridge metric consistently highlights these critical nodes, making the equation robust in finding candidates for targeted intervention in the misinformation network.

After optimizing the weights (w\_i, w\_e, w\_c) across the range of (1,1,1) to (10,10,10), we identified that the combination (10, 7, 7) produced the highest change in network density before and after perturbing the network, with 
\begin{math}
    \Delta\_density
\end{math}
being the maximum. The original network density was 0.0015849, which dropped to 0.0015668 after the removal of the top 12 bridge nodes.

This completed our bridge metric formulation and our proposed bridge metric is as follows:
\begin{equation}
\text{Bridge\_Score} = 10 \times \text{indegree\_centrality} + 7 \times \text{eigenvector\_centrality} + 7 \times \text{clustering\_coefficient}
\end{equation}
Using this equation, we compute bridge scores for all nodes in the network and order them from highest to lowest. Nodes with higher bridge scores serve as important cross-community hubs, facilitating rapid amplification of diverse narratives and reshaping network topologies. 
Our analysis identified a set of high-impact nodes that scored exceptionally well on our bridge metric, potentially serving as key conduits for cross-community information flow. 

\begin{table}[h!]
\centering
\begin{tabular}{|l|c|c|c|c|}
\hline
\textbf{Channel} & \textbf{Clustering Norm.} & \textbf{Indegree Norm.} & \textbf{Eigenvector Norm.} & \textbf{Bridge Score} \\
\hline
lomovkaa & 0.714286 & 1.0 & 1.0 & 22.000002 \\
lu\_di\_z & 0.8 & 0.714286 & 0.936702 & 19.299771 \\
warfakebelgorod & 1.0 & 0.571429 & 0.929098 & 19.217972 \\
zarussia\_1 & 0.533334 & 0.857143 & 0.546417 & 16.129686 \\
ua\_tribunal & 0.533334 & 0.857143 & 0.539694 & 16.082625 \\
zachitniki & 0.285714 & 1.0 & 0.547298 & 15.831084 \\
voin\_dv & 0.7 & 0.714286 & 0.532089 & 15.767480 \\
Lunay14 & 0.7 & 0.714286 & 0.532089 & 15.767480 \\
orchestra\_w & 0.7 & 0.714286 & 0.532089 & 15.767480 \\
rustroyka1945 & 0.7 & 0.714286 & 0.532089 & 15.767480 \\
cyber\_frontZ & 0.7 & 0.714286 & 0.532089 & 15.767480 \\
novosti\_efir & 0.7 & 0.714286 & 0.532089 & 15.767480 \\
heroesofZ & 0.466666 & 0.857143 & 0.546417 & 15.663010 \\
rtdru & 0.466666 & 0.857143 & 0.539694 & 15.615949 \\
\hline
\end{tabular}
\caption{Possible targets for intervention}
\label{tab:bridge_score}
\end{table}

Table 2. presents the top nodes ranked by bridge score, including their normalized values for in-degree centrality, eigenvector centrality, and clustering coefficient. This detailed breakdown provides insights into the diverse characteristics that contribute to a node's bridging potential.

\begin{figure}[h]
  \centering
  \includegraphics[width=0.65\linewidth]{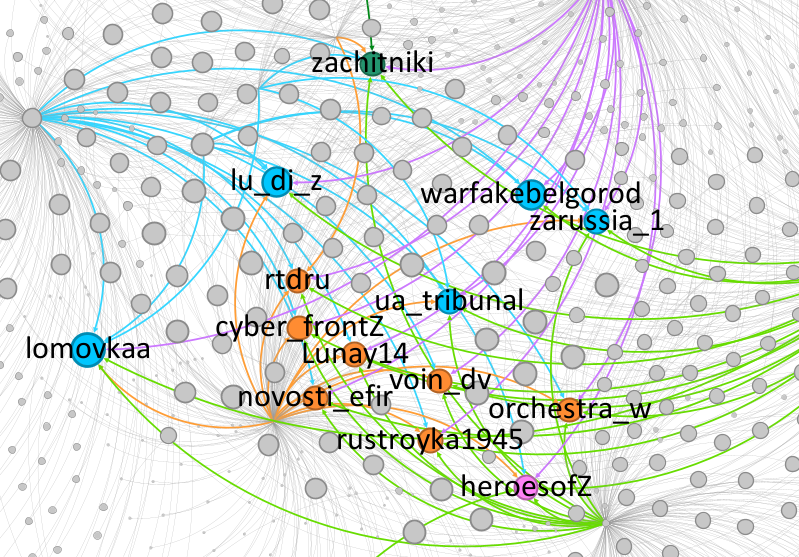}
  \caption{The most influential cross-community hubs}
  \Description{graph}
\end{figure}
\begin{figure}[h]
  \centering
  \includegraphics[width=1\linewidth]{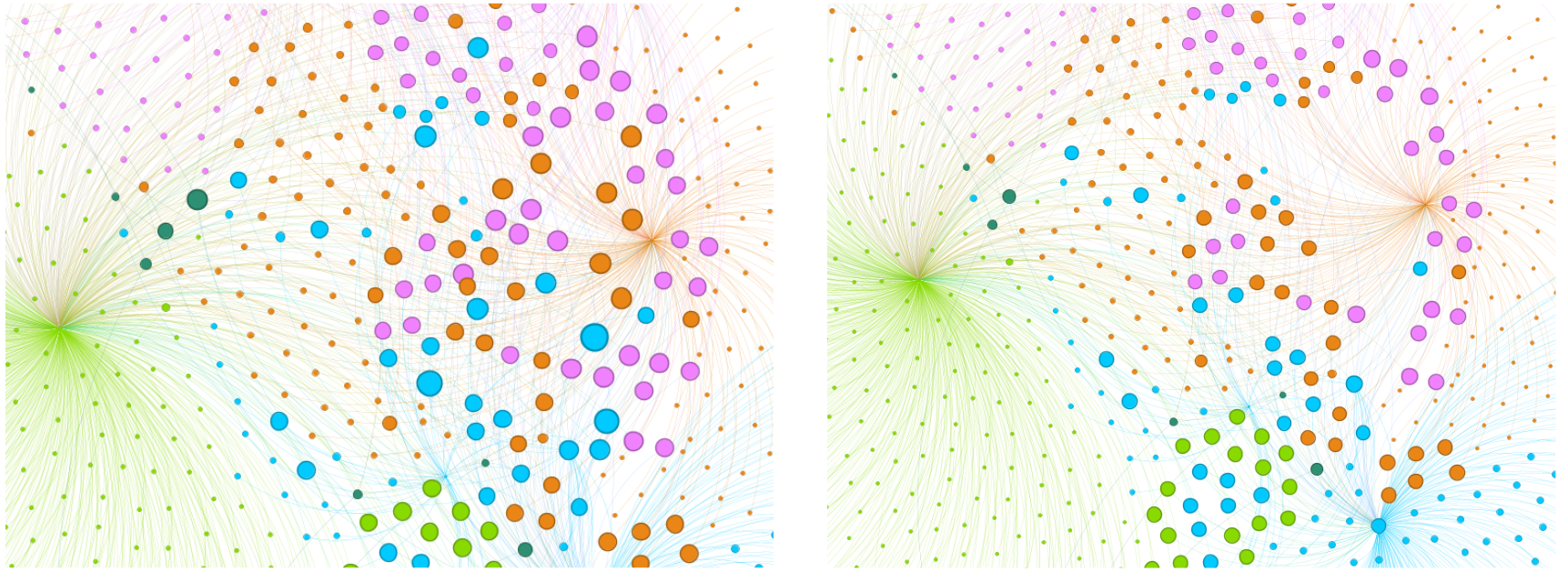}
  \caption{Network Visualizations Before and After Bridge Node Removal}
  \Description{graph}
\end{figure}

Figure 6 presents network visualizations before and after the removal of top bridge nodes identified by the optimal weight combinations (Figure 5), visually demonstrating their impact on network structure (Figure 7).

\subsection{Impact of Bridge Node Removal:}The removal of the top 12 bridge nodes identified by our optimal weight combinations resulted in:
\begin{itemize}
    \item 0.31\% increase in average path length (from 1.579 to 1.584). Even a small increase in path length suggests that removing key hubs slows down the spread of misinformation by increasing the distance between communities.

    \item 33.33\% increase in number of communities (from 6 to 8), highlighting how the removal of key bridge nodes fragments the network, splitting previously connected groups into smaller, isolated clusters. This fragmentation demonstrates the structural role these hubs play in binding diverse communities together, and their absence reveals the potential for misinformation networks to weaken, limiting the reach and cohesion of harmful narratives.

\end{itemize}

To evaluate the effectiveness of our proposed bridge score in reducing network density and fragmenting the misinformation network, we conducted a comparative analysis using nodes prioritized by the individual scoring metrics we used to build our bridge score: in-degree centrality, eigenvector centrality, clustering coefficient. In Table 3. each row presents the effects of removing the top nodes as ranked by a particular metric on key network properties—namely, average path length, number of communities, and network density.
By evaluating the impact of removing nodes ranked highly by individual scores—such as in-degree centrality, eigenvector centrality, and clustering coefficient—we observe that no single-metric approach achieves the same degree of network disruption as our composite bridge score.

\begin{table}[h]
\centering
\small
\begin{tabular}{|l|p{0.27\textwidth}|c|c|c|}
\hline
\textbf{Metric} & \textbf{Top bridge channels} & \textbf{Avg Path Length} & \textbf{\# Communities} & \textbf{Network density} \\
\hline
Indegree & lomovkaa, warfakes, zachitniki, dva\_majors, heroesofZ, rtdru, ua\_tribunal, zarussia\_1, cyber\_frontZ, denazi\_UA, IrinaVolk\_MVD, Kharkov\_Perviy, lu\_di\_z, Lunay14 & 1.216731 & 228 & 0.00175755 \\
\hline
Eigenvector & lomovkaa, lu\_di\_z, warfakebelgorod, zachitniki, warfakes, zarussia\_1, heroesofZ, dva\_majors, ua\_tribunal, rtdru, voin\_dv, Lunay14, orchestra\_w, rustroyka1945 & 1.216310 & 227 & 0.00175755 \\
\hline
Clustering coefficient & shkola\_voenkora, ves\_rf, infomil\_live, ZOV\_Voevoda, polk\_1430, domoy\_RF, zvofront, r\_vestovoi, Rubric\_lossesvsu, DnevnikDesantnika, TarasInform, AlabugaStartRussia, antiseptic\_channel, russian\_shock\_volunteer\_brigade & 1.573170 & 5 & 0.001591336 \\
\hline
Bridge score & heroesofZ, lunay14, novosti\_efir, orchestra\_w, rtdru, rustroyka1945, ua\_tribunal, voin\_dv, warfakebelgorod, zarussia\_1 & 1.579 $\rightarrow$ 1.584 & 6 $\rightarrow$ 8 & 0.001584 $\rightarrow$ 0.0015667 \\
\hline
\end{tabular}
\caption{Comparative Analysis of Network Disruption Metrics}
\label{tab:network-metrics-rows}
\end{table}

The impact of our work shows that by strategically targeting and removing such influential nodes, the overall network cohesion is severely weakened, disrupting the pathways through which misinformation spreads and compartmentalizing the spread of disinformation across distributed platforms.

\begin{figure}[h]
  \centering
  \includegraphics[width=0.8\linewidth]{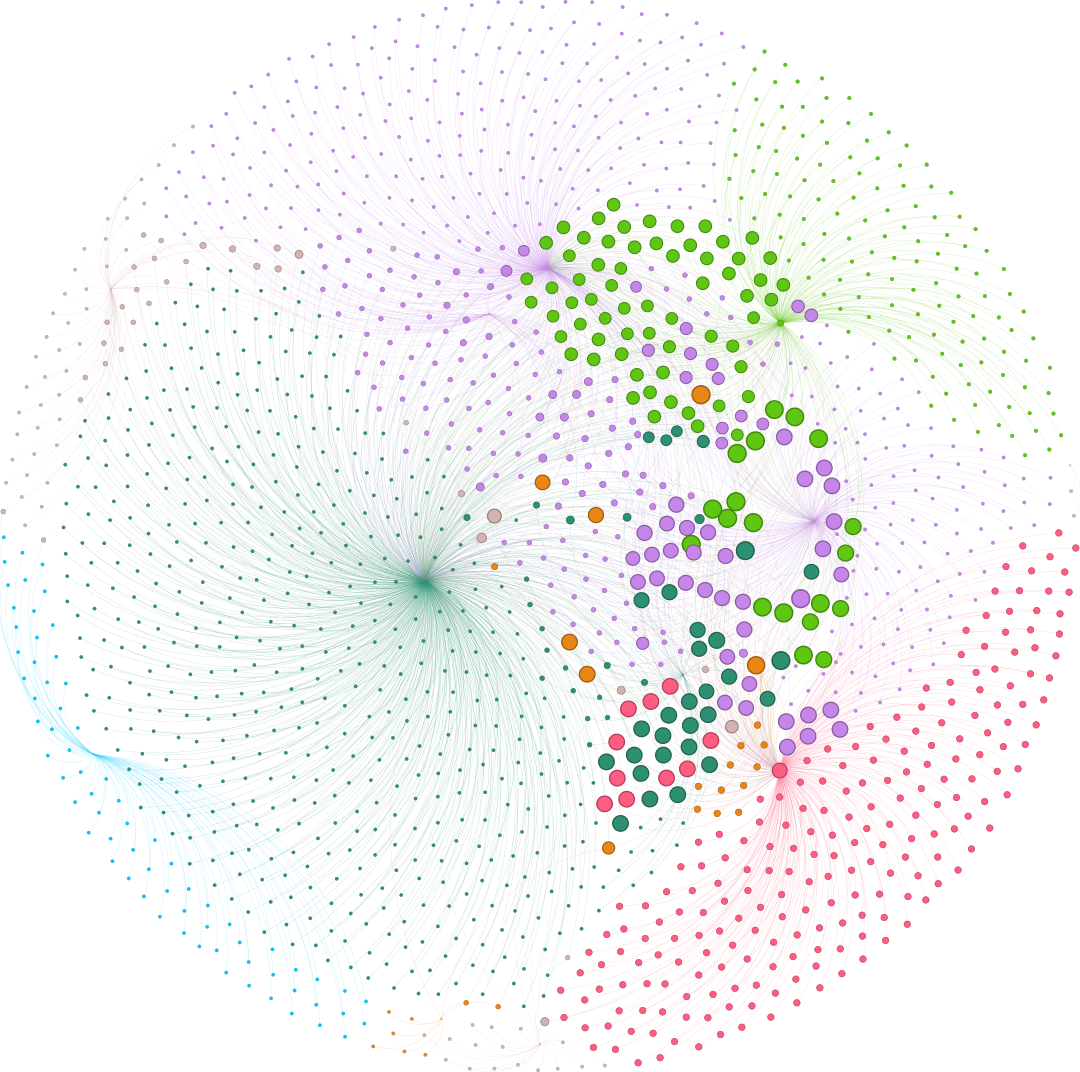}
  \caption{Distribution of network into 8 (originally 6) distinct communities  after removal of impactful hubs}
  \Description{graph}
\end{figure}

\subsection{Spatial Distribution of Bridge Nodes:}
Building upon the understanding of the network's structural characteristics, the bridge score formula is now applied  to identify and analyze key nodes that potentially facilitate cross-community information flow and misinformation propagation. To visualize the positioning of high-scoring bridge nodes within the network structure, we mapped their locations relative to detected communities.
\begin{figure}[h]
  \centering
  \includegraphics[width=0.8\linewidth]{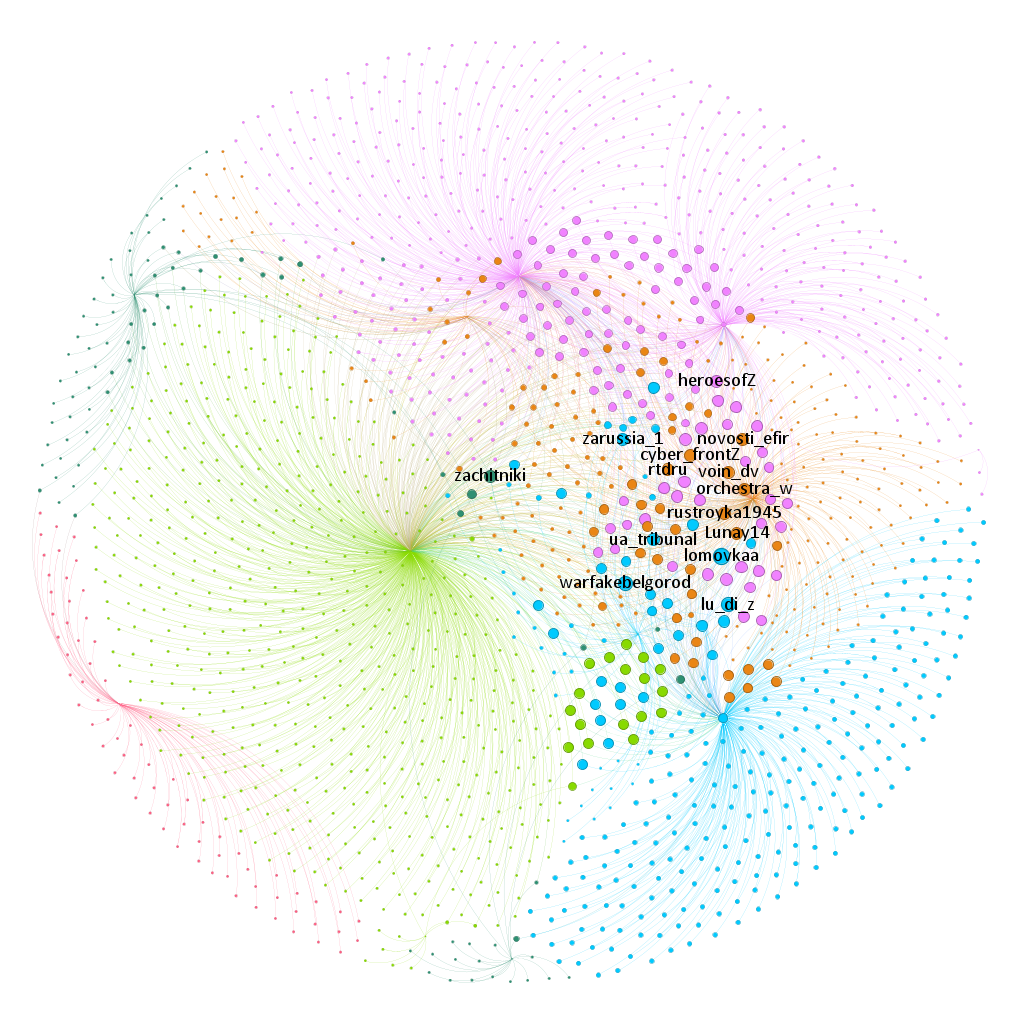}
  \caption{Central positions of bridge nodes in the network}
  \Description{graph}
\end{figure}

Figure 8 displays the network graph with nodes sized by their bridge score and colored by community membership. High-scoring bridge nodes are highlighted, revealing their tendency to occupy strategic positions at the center of the network and towards the internal peripheries of more prominent communities. With new channels being created frequently, there is a need for continuous monitoring. New hubs position themselves strategically to amplify misinformation of multiple community types.

\subsection{Significance of Cross-Community Hubs:}
The application of our multidimensional bridge score revealed a highly skewed distribution of node’s importance within the network.

\begin{figure}[h]
  \centering
  \includegraphics[width=0.9\linewidth]{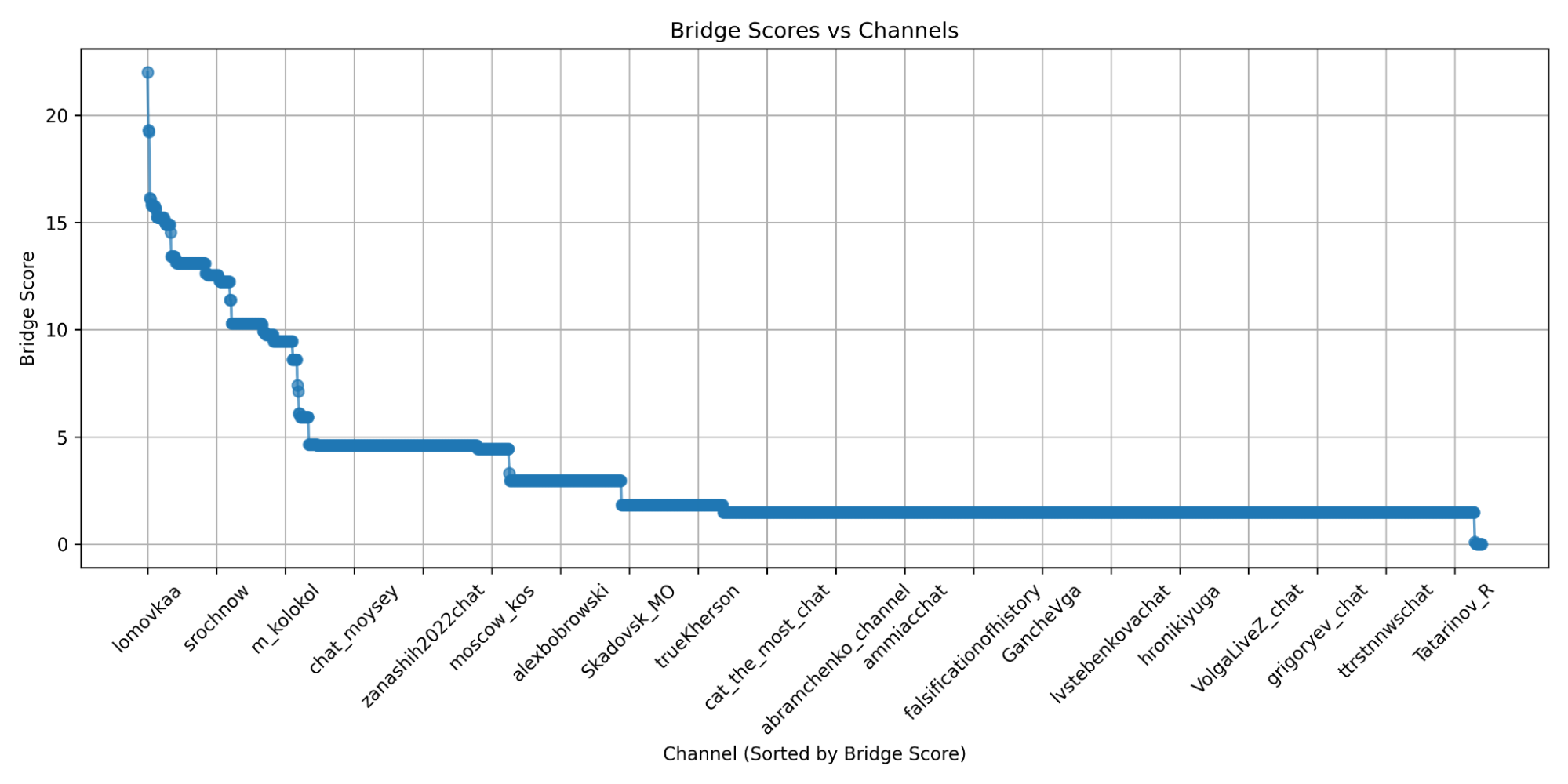}
  \caption{Bridge Score Log-log scale Distribution}
  \Description{histogram}
\end{figure}

Figure 9 presents a line graph of bridge scores across all nodes(with every 50th channel plotted across the X axis), plotted on a log scale. The distribution exhibits a heavy-tailed pattern, indicating that a small fraction of nodes possess disproportionately high bridge scores. This aligns with the scale-free nature of the network and suggests that a select few nodes play crucial roles in bridging communities.
A detailed analysis of engagement metrics between seed and bridge Telegram channels reveals striking contrasts. On average, messages in bridge channels are forwarded 845 times, significantly higher than the 147 forwards per message seen in seed channels. This indicates that bridge channels play a critical role in amplifying content, particularly misinformation, by distributing it to broader audiences. Additionally, bridge channels experience 3.1 times more views per message compared to seed channels, highlighting their widespread reach. Interestingly, the pattern for replies is reversed—seed channels receive around 34 replies for every 10 messages, compared to only 8 replies per 10 messages in bridge channels. This suggests that bridge channels limit interactive discussions, likely to prevent scrutiny or fact-checking, and focus on rapidly disseminating pre-existing narratives. Notably, most messages in bridge channels are forwards, further emphasizing their role in content amplification rather than original content creation.

\begin{figure}[h]
  \centering
  \includegraphics[width=0.95\linewidth]{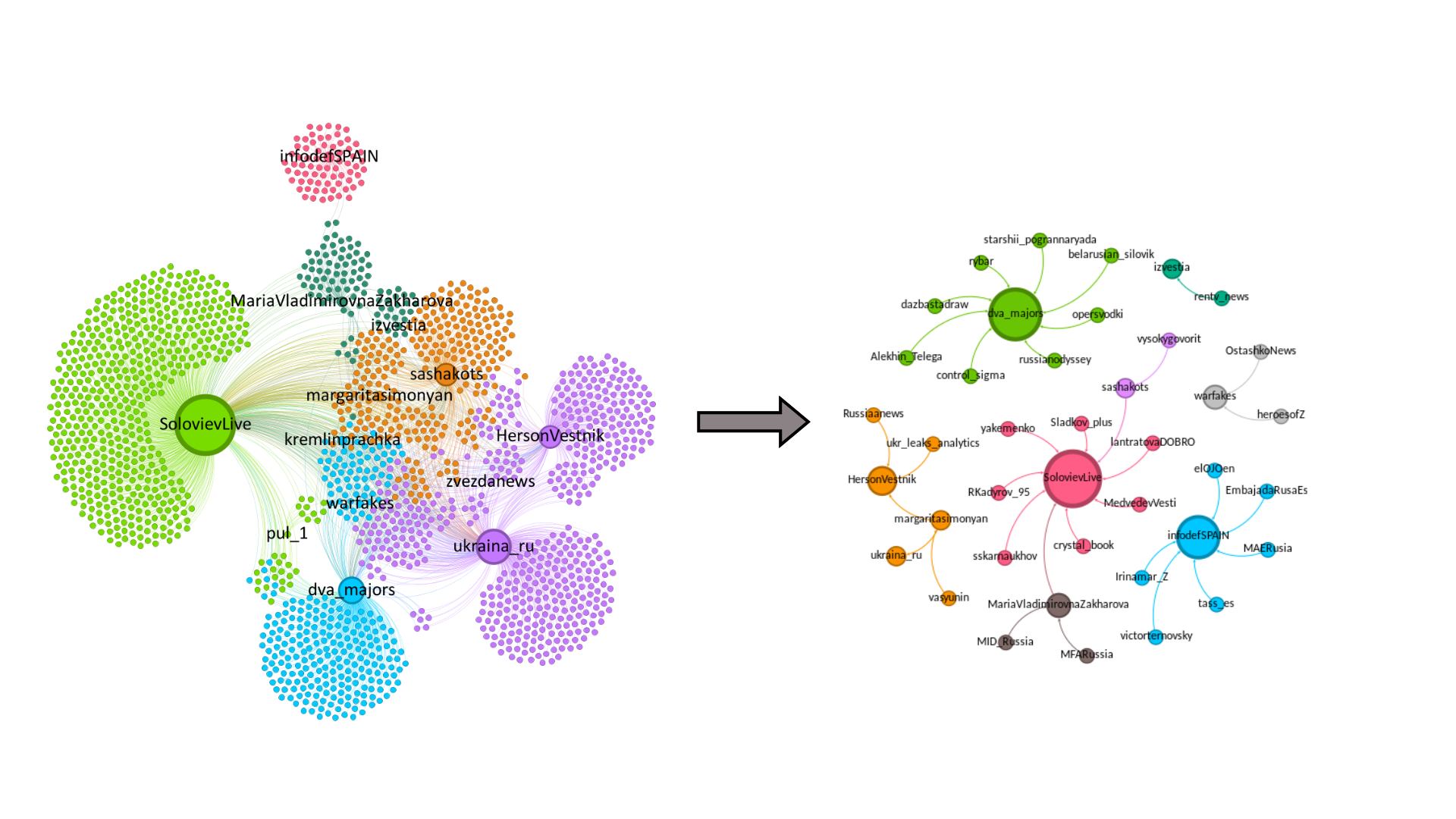}
  \caption{Russian disinformation machinery reuses the same Telegram channels to promote different campaigns}
  \Description{histogram}
\end{figure}

A close examination of the 4,492 Telegram channels that we originally started out with, reveal a recurring pattern in the Russian disinformation machinery, where the same network of channels is repurposed to promote various false narratives. Initially, these channels were mobilized to propagate disinformation surrounding the Russia-Ukraine conflict, spreading misleading narratives to large audiences. However, our analysis also shows that the same network was later reactivated to disseminate misinformation about the Moscow (Crocus Hall) attack as shown in Figure 10. This reuse of channels highlights the efficiency and adaptability of the disinformation ecosystem, where a pre-established network infrastructure is leveraged to quickly shift focus and promote different campaigns, depending on the geopolitical context. Such strategic mobilization of channels indicates a coordinated effort to maintain influence and control over public discourse, using misinformation as a tool to manipulate perceptions in various crises. This ability to recycle disinformation networks across different events underscores the importance of continuously monitoring these channels to counteract the evolving narratives they propagate.




A review of the links shared in bridging channels with corresponding url titles reveal that many of them lead to low-quality, unreliable \& state-sponsored sources, known for spreading misinformation and stoking social unrest. Examples include Dzen.ru \& Russian News Agency-TASS, who are Russian news aggregators, often cited for promoting state-aligned narratives and rated "Right-Center Biased" and "Questionable" by Media BiasFact Check\footnote{\url{https://mediabiasfactcheck.com/dzen-ru-bias/}, \url{https://mediabiasfactcheck.com/russian-news-agency-tass/}}. Notably, former intelligence agency personnel collaborated in this review to assess the nature and potential risks associated with these sources.

\begin{figure}[h]
  \centering
\includegraphics[width=1\linewidth]{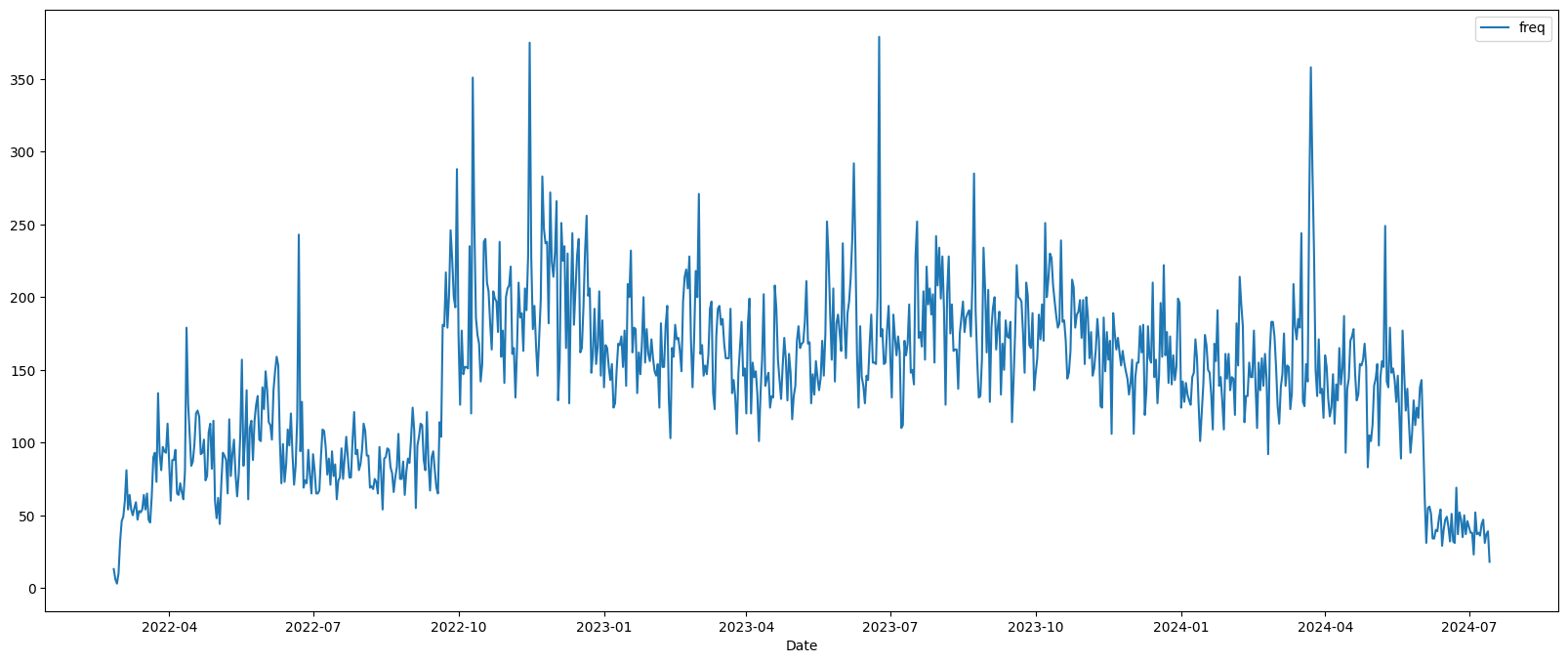}
  \caption{Message posting frequency in bridging channels with respect to time}
  \Description{histogram}
\end{figure}

Figure 11 shows an analysis of posting activity in bridging channels, which reveal a marked increase in frequency during major geopolitical events, often sparking controversy and potentially leading to social unrest. Notably, peaks were observed during events such as June 24, 2023, when the feud with Wagner Group owner Yevgeny Prigozhin posed a significant threat to Russian President Vladimir Putin’s 22-year rule, and October 13, 2022, when the U.N. General Assembly condemned Russia's illegal annexation in Ukraine. These spikes in activity indicate that such channels are strategically used to amplify divisive narratives during critical moments to destabilize public discourse. With time, their focus keeps changing and hence are capable of shaping a subscriber’s perspective with the misinformation relating to latest events and happenings.
By examining these nodes' strategic positions and stability, one can gain insights into how misinformation might spread across different groups. This information helps in developing targeted strategies to monitor and curb harmful content across network boundaries.

\section{Conclusion}
This work has contributed to the understanding of how distributed messaging platforms, particularly Telegram, facilitate the flow of information through and between communities. By focusing on the role of bridge nodes in cross-community information propagation, we demonstrated that these key nodes are instrumental in amplifying content across otherwise isolated clusters. Our introduction of multi-dimensional bridging metrics reveals the critical connection between network topology and the reach of information, showing that specific community structures can significantly enhance content visibility and engagement. Moreover, our temporal analysis of bridge node activity during significant geopolitical events highlights the dynamic and adaptive nature of information flow in distributed ecosystems. This underscores the importance of network structure in determining the conditions under which misinformation or other content achieves broader dissemination. The findings of this study extend beyond theoretical contributions by offering practical insights for platform design, content moderation, and intervention strategies. As misinformation continues to be a major challenge across digital ecosystems, this work provides a foundation for future research and the development of more robust socio-technical systems aimed at curbing the amplification of harmful narratives while promoting information integrity in distributed environments.


\begin{acks}
This research would not have been possible without the invaluable insights and expertise of Eric Brichetto and Roman Sannikov, whose work in open-source intelligence and disinformation investigations greatly enhanced the understanding of the different dynamics at play in misinformation campaigns. Their contributions have significantly shaped the direction and quality of this research.
\end{acks}

\bibliography{sample-base}

\appendix









\end{document}